\documentclass[prl,superscriptaddress,twocolumn]{revtex4}
\usepackage{graphicx}

\begin{document}

\title{Metamagnetic Quantum Criticality}
\date{\today}
\author{A. J. Millis}
\affiliation{Center for Materials Theory, Department of Physics, Rutgers
University, 136 Frelinghuysen Rd, Piscataway NJ 08854}
\author{A. J. Schofield}
\affiliation{School of Physics and Astronomy, University of Birmingham,
Edgbaston, Birmingham, B15 2TT, UK.}
\author{G. G. Lonzarich}
\affiliation{Cavendish Laboratory, Madingley Road, Cambridge, CB3 0HE, 
UK}
\author{S. A. Grigera}
\affiliation{School of Physics and Astronomy, University of St Andrews,
North Haugh, St Andrews, Fife, KY16 9SS, UK}

\begin{abstract}
A renormalization group treatment of {\it metamagnetic quantum criticality}
in metals is presented. In clean systems the universality class is found to
be of the overdamped, conserving (dynamical exponent $z=3$) Ising type.
Detailed results are obtained for the field and temperature dependence of
physical quantities including the differential susceptibility, resistivity
and specific heat near the transition. An application of the theory is made
to ${\rm Sr_{3}Ru_{2}O_{7}}$, which appears to exhibit a metamagnetic critical
end-point at a very low temperature and a field of order $5-7$T.
\end{abstract}

\maketitle

\begin{figure}[tb]
\includegraphics[width=\columnwidth]{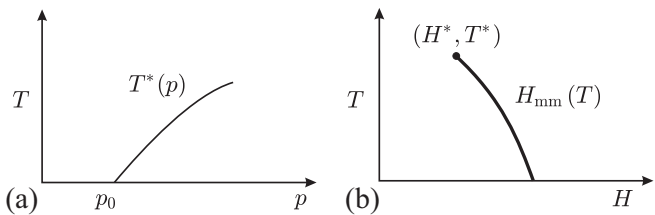}
\caption{(a) Schematic phase diagram, showing variation of end-point
of line of metamagnetic first order phase transitions as control
parameter (e.g pressure) is varied. (b) Schematic phase diagram in
$H,T$ plane for $p<p_0$ showing metamagnetic line and location of
end-point.}
\label{phasediag} 
\end{figure}

A metamagnetic transition is empirically defined as a rapid increase
in magnetization at a particular value of applied magnetic
field. Because there is no broken symmetry involved, one expects a
first order transition from a low magnetization to a high
magnetization state as an applied magnetic field $H$ is swept through
a (temperature dependent) critical value $H_{\rm mm}(T)$. The curve of
first order transitions $H_{\rm mm}(T)$ terminates in a critical point
$(H^{\ast },T^{\ast })$. By appropriately tuning material parameters
it is possible to reduce $T^{\ast}$ to $0$, yielding a {\it
quantum-critical end-point}.  This situation is depicted in
Fig.~\ref{phasediag}: (b) shows a typical metamagnetic line and
critical end-point in the field-temperature plane and (a) shows a
possible variation of the temperature of the critical end-point with
pressure.

A number of `strongly correlated metals', including 
${\rm UPt_{3}}$~\cite{UPt3ref}, ${\rm CeRh_{2}Si_2}$~\cite{CeRu2Si2ref}
and other heavy fermion compounds~\cite{metamagnetic}, 
as well as d-electron systems such as MnSi~\cite{Thessieu97} 
exhibit metamagnetic transitions with properties
suggestive of proximity to a quantum critical point. Recent measurements
strongly indicate that the bilayer ruthenate d-electron system 
${\rm Sr_{3}Ru_{2}O_{7}}$~\cite{Mackenzie00} is, 
at ambient pressure and moderate
applied field, tuned almost exactly to such a quantum-critical 
end-point~\cite{Grigera01}.

Metamagnetism in metals has not been extensively studied. Mean-field
treatments have recently appeared~\cite{Meyer01,Satoh01} and some
discussion was given in the context of a treatment of weak
ferromagnets via the `SCR' method~\cite{Yamada93}, but the critical
phenomena have not so far been investigated. Apart from the
experimental relevance, the issue is of fundamental importance. In a
metal at $T=0$ a nonzero magnetization corresponds to a splitting of
the ``spin up'' and ``spin down'' Fermi surfaces and a first-order
metamagnetic transition corresponds to a jump in this splitting. At
the metamagnetic quantum-critical end-point fluctuations of the
magnetization, and therefore of the Fermi surface positions, become
large: in a certain sense the material does not have a well defined
Fermi surface and may therefore be thought of as a
non-Fermi-liquid. In this paper we give what we believe is the first
renormalization group theory of metamagnetic quantum criticality. We
present general results, which may be applicable to a number of
systems, and a detailed application to ${\rm Sr_{3}Ru_{2}O_{7}}$. We use
the standard approach to metallic quantum
criticality~\cite{Hertz76,Millis93}, which involves integrating out
the electron degrees of freedom to obtain a model of overdamped
bosonic excitations which are analysed by renormalization group
methods.

Quotes are placed about ``spin up'' and ``spin down'' above because in many
metamagnetic materials spin orbit coupling is large and spin is not a good
quantum number. However, for most purposes one may adopt a `pseudo-spin'
notation~\cite{Blount85} labeling the two Kramers-degenerate states in
zero-field. The Kramers degeneracy is broken by an applied field, leading to
two Fermi surfaces and the theory carries through as in the non-spin-orbit
case with one important exception noted below: if spin-orbit coupling is
strong, impurity scattering may affect the dynamics differently.

To treat the critical behavior we assume that the fluctuations in the
relative Fermi surface positions may be represented by fluctuations of
a magnetization variable $m$. We write an action for longitudinal
fluctuations $\psi(x,\tau) =\left( \left| m\right| -m_{\rm av}(H^{\ast
})\right) /m_{0}$ of the magnetization density $m$ about its average
value $m_{\rm av}$ at the critical field $H^{\ast }$, normalized to
some typical magnetization density $m_{0}$ (for example, the
high-field saturation magnetization). We define the critical field
$H^{\ast }$ by the requirement that at $T=0$ the action has no static
third order terms. We write $h=\left( H-H^{\ast }\right) /H^{\ast }$,
introduce a cutoff length $a$ (for example, the lattice constant) and
define an energy scale $E_{\rm c}$ by the requirement that the
coefficient of the static quartic term is $1/4$; the action in $d$
space dimensions and imaginary time becomes
\begin{eqnarray}
S_{\rm meta}=S_{\rm dyn} + \quad \quad \quad \quad \quad \quad \quad
\quad \quad \quad \quad \quad \quad \quad &&
\nonumber \\ 
\int {d^dx \over a^d} E_c d\tau
\left[ \frac{1}{2}\xi_{0}^{2}
\left( \nabla \psi \right)^{2}+\delta \psi^2 + \frac{1}{4}
\psi^{4} 
 -h\psi + \dots \right] . && \label{smeta}
\end{eqnarray}
Here $\delta$ (which may be varied {\it e.g.} by changing
pressure) tunes the system through the metamagnetic critical point and
$S_{\rm dyn}$ (discussed below) expresses the order-parameter dynamics.

We have assumed in writing the static part of $S_{\rm meta}$ that the
gradient expansion is the conventional one, that the coefficients are
simple numbers and that the parameters vary with temperature only as
$T^{2}$, as usual in Fermi-liquid theory.  However, several recent
papers have called these assumptions into
question~\cite{Belitz97,Chitov01}. In particular, Belitz, Kirkpatrick
and Vojta~\cite{Belitz97} have presented perturbation-theory results
indicating that the expansion of the static spin susceptibility of a
conventional, non-critical clean Fermi-liquid about the $q=0$ value
contains terms of order $\left| q\right| $ (in $d=2$). Such terms
would invalidate the usual gradient expansion.  Subsequently, a
closely related $\left| T\right|$ temperature dependence of the
susceptibility and other parameters was studied~\cite{Chitov01}. The
available perturbative calculations~\cite{Belitz97} suggest that spin
rotational invariance is required to obtain the anomalous momentum
dependence, so that in the present situation the gradient expansion
might be the conventional one. However, one might expect $\xi_0^2$ to
be unusually large if $m_{\rm av}(H^\ast)$ is small.  A detailed
investigation of the behavior near a metamagnetic point could
therefore be a useful test of this still unsettled issue.

The dynamic part $S_{\rm dyn}$ follows because the order parameter is
essentially the difference in position of the spin-up and spin-down
Fermi surfaces. Fluctuations at nonzero $q$ correspond to locally
increasing the number of spin-up electrons and decreasing the number
of spin-down electrons. If spin is conserved such a fluctuation can
relax only via propagation or diffusion of electrons within each spin
manifold. In a clean spin-orbit-coupled system, pseudo-spin is
conserved (at least for fields aligned along a crystal symmetry axis)
and the same arguments apply. Therefore, in a clean system one expects
(the term is most conveniently written in frequency-momentum space)
\begin{equation}
S_{\rm dyn}=\frac{T}{E_{c}}\sum_{n}\int \frac{a^{d}d^{d}q}{\left( 2\pi \right)
^{d}}\frac{\left| \omega _{n}\right| }{v\left| q\right| }\left| \psi
(q,\omega _{n})\right| ^{2}+...  \; , \label{Sdyn}
\end{equation}
corresponding to overdamped but conserved fluctuations, and yielding
the dynamical exponent $z=3$. Here $v$ is a velocity, presumably of
the order of the Fermi velocity and $\psi (q,\omega _{n})=\int
\frac{d^{d}x}{a^{d}} E_{\rm c}d\tau e^{i\vec{q}\cdot \vec{x}-i\omega
_{n}\tau }\psi (x,\tau )$. Note that we are concerned only with
longitudinal fluctuations, so `precession' terms $\partial _{\tau
}\vec{\psi } \cdot \nabla^{2}\vec{\psi }\times \vec{\psi}$ are not
important. Strong (pseudo)-spin-conserving scattering would lead to
diffusion ($\left| \omega _{n}\right| /vq\rightarrow $ $\left| \omega
_{n}\right| /Dq^{2}$) changing $z$ to $4$. A momentum non-conserving
spin orbit coupling (as from impurities in the presence of strong spin
orbit scattering) would lead to relaxation ({\it i.e.} Eq.~\ref{Sdyn}
with $vq$ replaced by a momentum independent scattering rate) implying
$z=2$. An important scale is the characteristic energy $\omega _{\rm
sf}$ of a spin fluctuation at momentum $q_{\rm c}=1/a$, $\omega _{\rm
sf}=v\xi_{0}^{2}q_{\rm c}^{3}$.  Typical $\omega_{\rm sf}$ values for
transition metal magnets are of the order of 500K; for heavy fermion
systems they are at least an order of magnitude smaller~\cite{Params}.

We analyse the theory by the usual one loop renormalization group
equations~\cite{Millis93} which, after mode elimination and rescaling,
relates the theory with parameters $\delta,u,h$ to a new theory with
parameters $\delta ^{\prime },u^{\prime },h^{\prime }$. The behavior
at $h=0$ has been previously reported~\cite{Millis93}; we focus here
on the $h$ dependence.  The scaling equations are (we assume
henceforth that $z=3$)
\begin{eqnarray}
\frac{\partial \delta }{\partial \lambda } &=&2\delta +3u(\lambda
)f(T(\lambda ))  \label{delta} \; , \\
\frac{\partial u}{\partial \lambda } &=&\left( 1-d\right) u(\lambda )
\; . 
\label{u}
\end{eqnarray}
The field $h$ scales as $h(\lambda)=e^{{{d+5}\over 2}\lambda}$ and
$T(\lambda )=Te^{3\lambda }$. The effect of eliminated modes on
$\delta $ is contained in $f$ which is calculated by expanding the
theory about the value $\overline{\psi }(\delta h)$ which extremizes
the static part of $S_{\rm meta}$ at the rescaled field, and then
using the Gaussian approximation to the resulting action to evaluate
the integral over eliminated modes. Operationally, this means that we
calculate $f$ assuming the scales of interest are larger than the
running `mass' $r_{\rm eff}=\delta +3u\overline{\psi }^{2}$ and then
stop scaling at $r_{\rm eff}=1$. 
Expressing momenta and frequencies in units
of $q_{\rm c}=1/a$ and $\omega _{\rm sf}$ then 
$f=\Lambda \int^{^{\prime}}\frac{d^{d}u}{
\left( 2\pi \right) ^{d}}\frac{dy}{\pi }
\coth (\frac{y}{2t})
{y/u \over (y/u)^2+u^4}$ with $\Lambda =\left( \omega _{\rm sf}/
E_{\rm c}\right) \left(
a/\xi _{0}\right) ^{2}$ (the $'$ denotes summation over eliminated
modes). 

The solution of Eqs.~\ref{delta} and~\ref{u} follows~\cite{Millis93} and is
discussed in detail elsewhere~\cite{Millis01c}. Because in all cases of
physical interest the model is at or above its upper critical dimension,
quantal fluctuations lead only to a finite renormalization of the $T=0$
parameters of $S_{\rm meta}$ while thermal fluctuations are controlled by a
`dangerous irrelevant operator'. The mathematical consequence is that the
solution of the scaling equations may be written so the effects of quantal
fluctuations are absorbed into the $T=0$ parameters and only thermal effects
need be explicitly treated. We note, however, that the effects of quantal
fluctuations are not small in general, so that {\it a priori} one finds 
the parameters (such as $u$) of the theory are not well estimated by
band theory calculations.

The results of the calculation may be summarized as follows. At
$\delta =0$ ({\it i.e.} parameters tuned so that the material is at
the metamagnetic quantum critical point) as $T\rightarrow 0$ the
differential susceptibility $\partial m/\partial h$ scales as
$u^{1/3}h^{-2/3}$; as $T\rightarrow 0$ the specific heat coefficient
$\gamma =C/T$ is proportional to $\ln h^{-1}$ in $d=3$ and to
$h^{-1/3}$ in $d=2$ and the resistivity $\rho (T)$ has the leading $T$
dependence $\rho (T)-\rho (T=0)=AT^{2}$ with $A$ varying as $h^{-1/3}$
in $d=3$ and as $h^{-2/3}$ in $d=2$. The crossover to the thermally
dominated regime occurs at $T\sim h^{1/2}$ (d=3) and $T\sim h^{2/3} $
(d=2). If $\delta >0$ then the scaling in $h$ is cut-off when
$h^{2/3}\sim \delta$ and there is no phase transition in the $h,t$
plane. If $\delta <0$ then a first order transition occurs as $h$ is
varied at $T=0$; a line of first order transitions extends upwards in
the $h,T$ plane and terminates at a critical end-point temperature
$T^{\ast}\sim \delta ^{z/\left( d+z-2\right) }$. Finally, we note that
corrections to scaling may be numerically important, as will be seen
from the numerical results below.

The proceeding considerations were generic. It is possible to proceed
further in the particular case of ${\rm Sr_{3}Ru_{2}O_{7}}$, because it seems
(see below) that at ambient field this material is
very near to a weakly first order ferromagnetic-paramagnetic quantum phase
transition. The physics over a wide range of fields and temperatures should
therefore be describable by a generalized Ginzburg-Landau action for a
three-component order parameter $\phi $ (corresponding to long wavelength
fluctuations of the magnetization) 
\begin{eqnarray}
S_{0} =S_{\rm dyn}+  
\int \frac{d^{d}x}{a^{d}}d\tau \left\{ 
\frac{1}{2}\xi_{0}^{2}\left[\nabla_{b}\phi_{a}(x,\tau)\right]^{2}
+\frac{r}{2}\phi_{a}^{2}(x,\tau)\right. && \nonumber \\
+\left.
\frac{1}{4}u_{ab}\phi_{a}^{2}\phi_{b}^{2} 
+\frac{1}{6}v_{abc}\phi _{a}^{2}\phi_{b}^{2}\phi_{c}^{2}
-g_{\rm eff}\mu_{\rm B}\vec{h}\cdot \vec{\phi}(x,\tau )+ \dots \right\}.
&& 
\label{s0}
\end{eqnarray}
Here repeated indices are summed, the ellipsis denotes higher order
terms and the notations are as above, except that we have added a
sixth order term and here the parameters $r,u_{ab},v_{abc}$ have
dimension of energy. The data (isotropic susceptibility as
$T\rightarrow 0$ but some angle dependence in high-field magnetization
and higher temperature $\chi$) require a breaking of rotational
invariance in the mode coupling terms, but not in the quadratic
one. We take $\phi$ to be a dimensionless magnetization variable
measured in units of the putative saturation magnetization
$2\mu_{B}/{\rm Ru}$ (the important electrons are $d$ electrons, of
which there are four in the $t_{2g}$ orbitals, leaving two holes, and
the $g$ factor should be close to $2$). Scaling is as described
previously, except the sixth order term renormalizes the fourth order
one and in the presence of a field the `mass' (coefficient of the
quadratic part of the fluctuations) becomes anisotropic, with the
component corresponding to fluctuations along the field becoming
$r_{\rm eff}=r+3u\overline{\phi }^{2}+5v\overline{\phi }^{4}$.  A
Heisenberg-$XY$ or Heisenberg-Ising crossover occurs when then larger
of $r_{\rm eff}$ or $r$ passes through unity and scaling stops when the
smaller of the two becomes of order unity. In the Heisenberg regime
extra `precession' terms in the dynamics may be important. A detailed
analysis of the behavior of this model will be presented
elsewhere~\cite{Millis01c}, extending the important work of Yamada and
collaborators~\cite{Yamada93}, who showed that such an analysis was
possible but did not consider the precession terms or anisotropic
scaling and also used a simplified version of the `SCR' theory instead
of the renormalization group method. At $T=0$, one may use mean-field
theory as before, provided one interprets the parameters in
Eq.~\ref{s0} as renormalized parameters. If $9u^{2}/20v>r>0$ and $u<0$
(for simplicity we do not write the directional subscripts here) the
model has a $T=0$ metamagnetic transition. The point $r=9u^{2}/20v$
corresponds to the quantum critical end-point. At the quantum-critical
end-point the magnetization $m=\langle\dot{\phi}\rangle$ and magnetic
field $H^\ast$ are, for fields in the c-direction,
\begin{equation}
m_{c}^{\ast}=\sqrt{\frac{-3u_{cc}}{10v_{ccc}}} \; , 
\quad \quad
g_{\rm eff}\mu_{\rm B}H_{c}^{\ast}=
\sqrt{\frac{-3u_{cc}}{10v_{ccc}}}\frac{6u_{cc}^{2}}{25v_{ccc}} \; .
\label{endpt}
\end{equation}

Fig.~1 of Ref.~\cite{Mackenzie00} shows that at low $T$ and low
applied field the susceptibility is about $0.025\mu_{\rm B}/T$
implying $r\approx 160\mu_{\rm B}-T\approx 100$K.  This small value
implies a very large enhancement of the susceptibility over the band
value, as noted previously, and implies that the material is near a
paramagnetic-ferromagnetic transition. For fields directed along the
$c$ axis the observed metamagnetic transition occurs at a
magnetization of about $0.25-0.3\mu_{\rm B}/{\rm Ru}$ implying
$u_{cc}=3000-4300$K and $v_{ccc}=40,000-80,000$K with the larger
values corresponding to the smaller $m$. The consistency of these
estimates may be verified by substitution into Eq.~\ref{endpt}; use of
$g_{\rm eff}=2$ yields an estimate of $5-6$T for the metamagnetic
field, in the range found experimentally.  Expansion of Eq.~\ref{s0}
about the metamagnetic point yields Eq.~\ref{smeta} with $E_{\rm
c}=-2u_{cc}=6000-8000$K. The dimensionless critical field
$g\mu_{B}H^{\ast }/u\sim 0.001$ so we should be concerned with
variations which are small relative to this, {\it i.e.}  with
$h\approx 10^{-4}$. At present rather less information about the spin
fluctuation frequencies is available; we therefore normalize our
results to the temperature $T_{0}$ at which the differential
susceptibility at the critical field is equal to the zero-field zero
temperature susceptibility, {\it i.e} $\frac{\partial m}{\partial
h}(\delta =0,T=T_{0})=\chi(H=0,T=0\dot{)}=\chi_0$.

\begin{figure}[tb]
\includegraphics[width=\columnwidth]{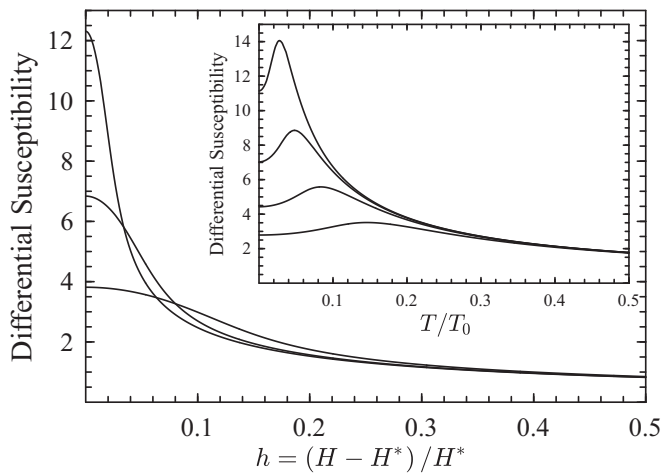}
\caption{Differential susceptibility, $\chi_0^{-1}(\partial m/\partial
h)$, as a function of applied field $H$ at temperatures
$T/T_{0}=0.05,0.1,0.2$, for a two dimensional metamagnetic critical
point.  Inset: Dependence of $\chi_0^{-1}(\partial m/\partial h)$ on
temperature $T$ at $h=.01,.02,.04,.08$. (Normalizations discussed
in text.)}
\label{chi} 
\end{figure}
\begin{figure}[tb]
\includegraphics[width=\columnwidth]{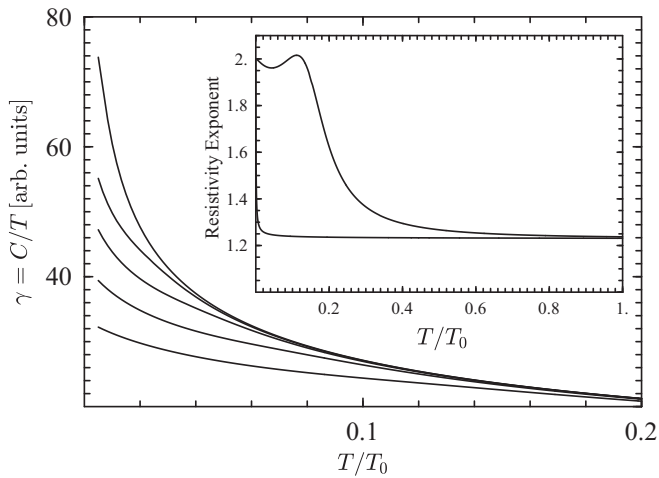}
\caption{Dependence of specific heat coefficient $C/T$ on temperature
$T$ for $h=0.01,.0.1,0.2,0.4$ calculated for a two dimensional
metamagnetic critical point.  Inset: Dependence of resistivity
exponent $\partial \ln \rho /\partial \ln T$ on $T$ for $h/H^{\ast
}=0$ (lower curve) and $0.1$ (upper curve). ($T_0$ defined in
text.)}
\label{gammarho} 
\end{figure}

We now present the results of a numerical solution of the scaling
equations.  Fig.~\ref{chi} shows the $h$ dependence of the
differential susceptibility for several values of $T$,
obtained in the two dimensional case using parameters reasonable for
${\rm Sr_{3}Ru_{2}O_{7}}$.  The inset shows the temperature dependence
of the differential susceptibility for different $h$. Note the
non-monotonic temperature dependence for fields different from $h=0$
if the control parameter is tuned to criticality.  Fig.~\ref{gammarho}
shows the specific heat coefficient $\gamma =C/T$; in this quantity
the crossover is much less sharp, in part because a 2d nearly critical
Fermi liquid has a specific heat coefficient $\gamma \sim A+BT$ with
both $A$ and $B$ divergent as the critical point is approached. This
is an example of the corrections to scaling mentioned earlier. The
inset shows the resistivity exponent $\alpha =-\partial \ln \rho
/\partial \ln T$ plotted against temperature for $h=0$ and
$h=0.1$. The high-$T$ resistivity exponent is not precisely $4/3$
because of the logarithmic corrections alluded to earlier. The
crossover to the expected low-T $T^{2}$ behavior is very sharp.

In conclusion, we have presented a theory of metamagnetic quantum
criticality in metals, which should be amenable to detailed
experimental tests. The universality class was identified, a form for
the order parameter dynamics was obtained, and detailed results were
presented for a range of physical quantities. Subsequent
papers~\cite{Millis01c} will present details omitted here as well as
results for three dimensional materials, and a comprehensive analysis
of ${\rm Sr_3Ru_2O_7}$ and MnSi.  The key assumption is that
electronic degrees of freedom may be integrated out, and the
fluctuating Fermi surface expressed in terms of an overdamped bosonic
magnetization variable. Experimental tests of the theory will show
whether this assumption, which underlies much recent work on quantum
phase transitions in metals and indeed on non-Fermi-liquid physics, is
justified.

{\it Acknowledgments}: AJM was supported by NSF-DMR-0081075 and the
EPSRC and thanks the Cavendish Laboratory, the Theoretical Physics
group at the University of Birmingham and the Aspen Center for Physics
for hospitality while parts of this work were undertaken. AJS
acknowledges the support of the Royal Society and Leverhulme Trust. We
thank A. P. Mackenzie for advice, encouragement and comments on the
manuscript.

\end{document}